# Low-Energy Sensor Network Time Synchronization as an Emergent Property

*Stephen F. Bush, Senior Member, IEEE*

GE Global Research Center, Niskayuna, NY, 12309, USA

## ABSTRACT

The primary contribution of this work is to examine the energy efficiency of pulse coupled oscillation for time synchronization in a realistic wireless network environment and to explore the impact of mobility on convergence rate. Energy coupled oscillation is susceptible to interference; this approach uses reception and decoding of short packet bursts to eliminate this problem. The energy efficiency of a commonly used timestamp broadcast algorithm is compared and contrasted with pulse-coupled oscillation. The emergent pulse coupled oscillation technique shows greater energy efficiency as well as robustness with mobility. A proportion of the sensors may be integrated with GPS receivers in order to obtain a master clock time.

## 1. INTRODUCTION

Time synchronization is a critical component of sensor networking. Sensors are becoming smaller and the data they collect yield higher resolution; higher resolution timing will be required to correlate finer-grained data. Time synchronization is also critical to sensor network efficiency and performance; any technique to improve sensor network efficiency by powering down transmitters and receivers will be limited by the accuracy of time synchronization.

The time synchronization protocol itself consumes energy; the contribution of this paper is to examine the energy efficiency of a variant of Pulse Coupled Oscillation (PCO) [4] using an energy model that takes into account the power required for frequent circuit startup and shutdown as well as receiver and transmitter power. This is primarily due to the Local Oscillator (LO) whose energy and spin up time is mitigated as discussed later.

A widely used technique for time synchronization is the Network Time Protocol (NTP) [6]. A server, or a well-organized hierarchy of servers, transmits messages containing current time to a registered set of client clocks. Variations of the NTP allow for greater accuracy by estimating and reducing sources of variability in message transfer and processing times. While NTP yields reasonable results in wired networks, a finer degree of accuracy will be required for wireless sensor networks due to tighter interaction between sensors and their environment.

Recent work on-time synchronization for sensors includes Reference Broadcast Synchronization (RBS) [1]. RBS begins to move away from an absolute time, instead it synchronizes to a relative event, namely the receive time of a broadcast signal among a group of receivers. It is assumed that transmission-time is fast enough that all receivers receive a broadcast transmission instantaneously. This serves as a point of reference for time synchronization. RBS does not scale to distances in which variance in propagation delay approaches the desired synchronization accuracy, however, for most terrestrial applications the assumption of instantaneous reception may be reasonable. In RBS, receivers must exchange times of reception of the reference signal in order to determine relative synchronization error. Thus, there is a message containing a time-stamp that must be exchanged; the number of exchanges grows with the number of receivers. With this information, any node's clock can synchronize to the clock of any other node with whom it exchanged a time-stamp.

The various flavors of IEEE 802.11 [5] time synchronization are similar to NTP in that a single timeserver is used to synchronize a set of clocks by periodic transmission of an absolute time stamp. Time synchronization messages must compete with IEEE 802.11 traffic and are subject to collision. Time synchronization in 802.11 is not scalable. Energy consumption is relatively high for 802.11 time synchronization because a single timeserver must enforce its notion of time by transmissions powerful enough to reach all receivers.

Synchronization energy consumption is sensitive to message size, rate of message exchanges, and RF attenuation of message transmission, based on distance as shown in Fig. 1. Synchronization algorithms trade-off along dimensions in this space as attempts are made to minimize energy and maximize synchronization performance. Points in this three-dimensional space have not been directly measured, but rather serve as an illustration of the tradeoffs. A single centralized time synchronization beacon, which transmits a timestamp to all nodes, must reach all nodes for successful synchronization. Thus it resides on the low rate, long distance, and relatively large size dimensions.

RBS requires a broadcast to all nodes requiring synchronization; multiple broadcasts are required to improve time dispersion. Additional messages are required to synchronize the receivers. This places RBS at a moderate message size, low rate, and moderate transmission distance position. NTP requires the transmission of timestamps while attempting to compute variances in link latencies (geared towards wired links). NTP is thus categorized in the large message size, short transmission distance, and medium rate position. IEEE 802.11 synchronization uses a relatively large message size, higher rates, and moderate transmission distances.

A subspace of the volume shown in Fig. 1 that has not been explored as fully as the others resides in the higher rate, shorter distance, and smaller message size dimensions. Higher rates of message exchange would not, *prima facia*, sound intuitively appealing as a low energy solution, however, this paper examines the use of periodic high rate bursts to achieve synchronization with low energy. The emergent technique is located in the small message size, short transmission distance, and high rate area of the figure below.

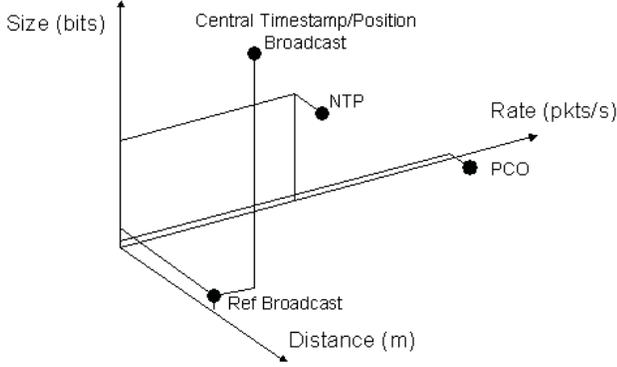

Fig. 1. Qualitative tradeoffs impacting energy consumption: message size, message exchange rate, and transmission distance.

A discussion of a low-power emergent technique for sensor time synchronization follows in Section 2. Experimental validation via simulation of emergent synchronization is presented in Section 3. Section 4 discusses the impact of mobility on the sensor network and presents experimental validation simulation results with node mobility.

## 2. EMERGENT TIME SYNCHRONIZATION: PULSE COUPLED OSCILLATION (PCO)

The feasibility of an emergent technique based upon on pulse-coupled oscillators [3] is explored as a low power solution to time synchronization. Equations describing the interaction of coupled oscillators have been developed in previous work [7]; this paper briefly describes PCO for sensor network time synchronization. The PCO algorithm uses a leaky integrate-and-fire mechanism based upon a state variable, $x_i(t)$. The state variable takes values in the range $[0, x_{th}]$, where $x_{th}$ is a threshold value. When $x \geq x_{th}$, a pulse is emitted. Integration occurs over the difference between an internal excitation variable, $S_0$, and leakage in the form of $\gamma x_i$. Thus, $x_i$ varies according to $\frac{dx_i}{dt} = S_0 - \gamma x_i$. The pulses emitted are observed and the time difference between the last pulse $t_{p-1}$ and current pulse $t_p$ are measured and normalized by the time interval $T_d$ towards which convergence is desired yielding $\phi_i(t) \in [0,1]$ in (1). The desired synchronization is achieved when $\phi_i(t) = 1.0$ for all nodes.

$$\phi_i(t) = \frac{t_p - t_{p-1}}{T_d} \quad (1)$$

The pulses emitted are assumed to have inter-pulse durations much smaller than the desired convergence interval. Each node in the system emits pulses that are received by a subset of nodes in the network. The signal received provides a coupling energy that is added to the original equation yielding $\frac{dx_i}{dt} = S_0 - \gamma x_i + \xi(r_i(t))$ where $r_i(t)$ is the received signal at node $i$ from all other nodes at time $t$. It is well known that such systems will converge to a common frequency [7]. This technique is emergent because each node makes a local decision based upon information from nearest-neighbors that yields a common global response, namely convergence to a common pulse rate. Local detection of convergence occurs when the phase change is constant. This indicates that all nodes have reached consensus on a common event, namely, the last pulse in which no further phase shift has occurred. Local time is adjusted on each node to match this synchronous pulse.

### 2.1. K-Nearest Neighbor Power Control

Consider the case a centralized broadcast to perform synchronization. If the distance between node $i$ and $j$ is $d_{ij}$ then, even in the best case, any broadcast approach would require a transmission distance of $\max_{j \in Nodes} d_{cj}$ where $c$ is the node positioned closest to the center of the sensor location distribution.

If 1-nearest neighbor PCO is utilized, the total network transmission distance is $\sum_{i,j \in Nodes} \min_d d_{ij}$. Comparing the centralized beacon and the 1-nearest neighbor PCO technique, the transmission distance of the 1-nearest neighbor is always equal or smaller than the centralized broadcast, $\max_{j \in Nodes} d_{cj} \geq \sum_{i,j \in Nodes} \min_d d_{ij}$ where $\min_d d_{ij}$ indicates that node $i$ and node $j$ are nearest neighbor pairs.

Generalizing to a uniformly random distribution of $n$ sensors over a circular area $A$, the node density is $\frac{n}{A}$, and the expected nearest-neighbor distance is $\sqrt{\frac{A}{n}}$. Computing the average distance between nodes and adding this potential error to the radius determines the expected distance of a node from the precise center of the node distribution. The expected distance between nodes is $\sqrt{\frac{A}{n+1}}$. In general, the sum of the nearest-neighbor distances squared represents free space path loss for one transmission and is shown in (2) where $\delta$ is the attenuation exponent and is assumed to be 2.

$$\sum_n \left(\sqrt{\frac{A}{n}}\right)^\delta \quad (2)$$

In a similar manner, the broadcast distance pathloss is shown in (3) where the first term is the radius of the node distribution area and the second term is the expected distance a node will actually lie from the exact center. A single beacon must cover the entire area in (3) and the further the beacon is from the center of the node distribution, the longer the transmission range.

$$\left(\sqrt{\frac{A}{\pi}} + \sqrt{\frac{A}{n+1}}\right)^\delta \quad (3)$$

The ratio of (3) over (2) indicates the gain in 1-nearest neighbor transmission distance of PCO per transmission over an optimally located central beacon. As density increases, the gain rises as shown in Fig. 2. Clearly, power reduction becomes more significant with a larger density of nodes.

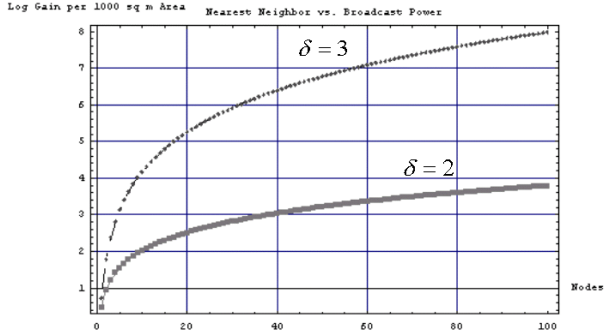

Fig. 2. Gain in power reduction versus number of nodes using nearest neighbors with pathloss exponents of two and three.

**2.2. Power Consumption Model**

A single time-stamp broadcast message of 180 bits with a transmission rate of 4 Mb/second requires that the receiver remain on for 0.045 milliseconds compared with a 5-millisecond window for PCO in which very short pulses are used. Each PCO pulse lasts only 0.2 milliseconds within the 5-millisecond window. The 5-ms window is used because the on/off transition rate is limited by the local oscillator (LO) warm-up time. The 5-ms window reduces the circuit on/off transition rate and the corresponding startup power consumption. The LO warm-up time is on the order of 450 microseconds, which is a limiting factor in the ability to transition between on and off states rapidly. Thus, in the PCO case, it is assumed that the LO remains on for a 5 mS time period to achieve synchronization. The LO startup power as well as receiver power are included in the energy model of the simulation described later in the paper. Note that PCO messages are decoded in order to avoid corruption by noise or malicious energy sources, however, PCO messages contain no explicit payload.

GPS enabled nodes may serve as master clocks for the non-GPS nodes. Master clocks will participate in the pulse-coupled oscillations by emitting pulses at the desired frequency. However, master nodes (nodes containing master GPS clocks) will not integrate received pulses or change their pulse frequency. Thus, master nodes will serve to drive non-GPS nodes, which are integrating and adjusting inter-pulse times according to the PCO algorithm. The 0.2-millisecond pulse window is derived from an assumption of 16 bits per pulse at 4 Mbs and a conservative estimate of 50 oscillations to reach convergence. The time synchronization duty cycle is thus $10^{-5}$.

The algorithm performs a summation of the number of short pulse packet arrivals rather than directly integrating energy. This induces the system to generate a series of tight pulse packet inter-arrival times. Fig. 3 illustrates a PCO time synchronization duty cycle. Clocks are assumed to drift at a rate of $10^{-8}$ seconds. The receiver could remain on to integrate over PCO pulses for five milliseconds every 500 seconds in order to maintain a +/-five microsecond clock accuracy. The convergence accuracy of the pulse coupled oscillator system depends on nearest-neighbor distances, $\sum_d \min d_{ij}^2$. Maximum oscillation frequency is limited by propagation delay of $\frac{\min_d d_{ij}}{c}$ where $c$ is the speed of light. If we assume an expected distance of 300 meters between sensors, then one microsecond error is possible. However, this is well within the refractory period of our PCO simulation model and did not appear to be a significant factor in the simulation results described later. If we assume a 50 mW requirement for transmit and receive power, then the total receiver energy is $50mW \times 5mS = 250\mu J$. We anticipate building low-power GPS receivers requiring one mJ or less per synchronization update.

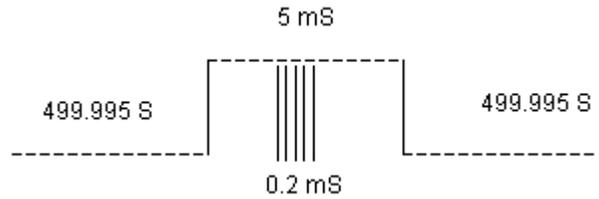

Fig. 3. The synchronization duty cycle is comprised of a series of converging pulses within a 5 ms receiver on time window.

**3. EMERGENT SYNCHRONIZATION: EXPERIMENTAL VALIDATION**

The energy model implemented in the simulation whose results are described in this section accounts for circuit startup power consumption, transmission power, and receiver power using CSIM [8], a general-purpose discrete-event simulator for block diagram oriented systems, on a GNU/Linux 2.4.20-20.9 Operating System. Accurate timing and energy consumption for 4 Mbs transmission rate has been implemented in the simulation with a local oscillator startup time of 450 milliseconds. Virtual clocks are implemented to simulate clock drift, which is currently $10^{-8}$.

The emergent clock synchronization technique using pulse-coupled oscillators is implemented with packet sizes of 16 bits, while the GPS master clock node broadcasts a timestamp packet of 180 bits. In this particular simulation, there is a single randomly chosen GPS node. Future experiments will consider multiple GPS nodes and the impact of their distribution within the PCO network. The PCO threshold in this simulation determines when internal excitation is large enough to cause a pulse to fire and is set to 3. In this particular simulation scenario, there are 612 randomly located sensors in a one-kilometer square area. In the initial experiments, the sensors are statically located, in a following set of experiments, all sensors move according to a Brownian motion movement model.

The clock phase differences, which are held constant for central broadcast and is variable for PCO (phi) is also sampled as are the number of pulses and time to reach synchronization (TTS). A power control mechanism is used in the emergent technique in which power is gradually increased until a response from a nearest neighbor is received. This sets the PCO pulse power, but does not guarantee a fully connected network of sensors.

The propagation model of Hata-Okumura (HO) [2] is used to account for signal attenuation. The HO parameters are a frequency of $10^9$ Hz, a rural country environment, and transmit and receive antenna height are 1 meter. A best-case scenario is used for the broadcast technique; a geographic center-most node broadcasts a time stamp to all nodes.

The total synchronization power consumed over the duration of the simulation indicated the emergent technique used approximately 15 times less power than the centralized timestamp broadcast. As previously mentioned, this includes circuit startup costs, as well as transmission and reception power consumption. Clearly, the longer transmission distance and larger message size of broadcast techniques dominates smaller shorter transmissions of the PCO technique. Assuming receiver power consumption is 50 mW, it is clear that receiver power has a greater impact on total power consumption since PCO total power consumption is much lower than the alternative technique. The timestamp broadcast algorithm chooses its power to be optimal for a centralized broadcast, since this is the lowest possible power requirement for the broadcast case.

Let $s$ be the number of nodes that are synchronized to a pre-specified tolerance and $n$ be the total number of nodes. $s/n$ is the proportion of nodes that are synchronized. A suggested metric is the proportion of nodes that remain unsynchronized, $1 - s/n$, which will be referred to as the proportion-out-of-sync (PoS). Synchronization efficiency, defined as the proportion of nodes within synchronization tolerance normalized by the total power of both transmitter and receiver is shown in Fig. 4. PCO appears to display greater power efficiency than timestamp message broadcast.

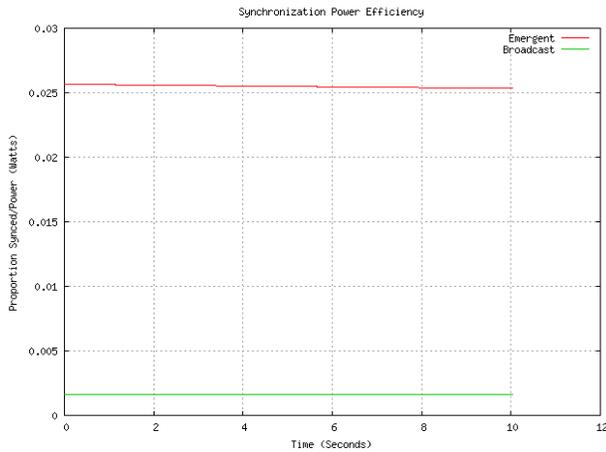

Fig. 4. Synchronization efficiency is the proportion of nodes synchronized (PoS) normalized by power.

The longer it takes for a system to synchronize, the more energy is consumed and the longer it must operate out of synchronization, which causes the system to operate in a higher energy state, e.g. more collisions, longer guard-bands, higher probability of misrouting, et cetera. Time to Synchronize (TTS), in units of milliseconds measures the time to reach synchronization for each round of synchronization.

The expected time to reach synchronization is 0.01 seconds with a variance of 0.0002 seconds[2] given the current PCO configuration. Details on time to reach synchronization are shown in Fig. 5. The initial spike in the time to reach synchronization is due to the initial power control algorithm that adjusts power sufficient to reach the nearest neighbor. In the static model used here, node density enables a fully connected graph using 1-nearest neighbor. If 1-nearest neighbor does not result in a connected topology, then islands of synchronized nodes will form. If enough GPS enabled anchor nodes with line of sight to a satellite are distributed throughout the network, 1-nearest neighbor may still yield synchronization, however, this scenario will be explored in a follow-up paper.

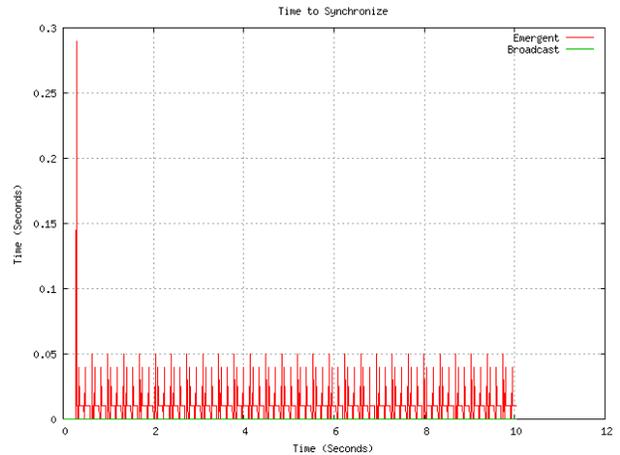

Fig. 5. TTS is an average taken from the last plotted time to the current time. The first synchronization time period is long due the power-control algorithm employed to discover nearest neighbors.

Expected transmission delays are measured only for actual transmissions. Thus, the expected values and variances are dependent upon which nodes actually fire in the PCO algorithm. The variance in clock time is plotted in Fig. 6. In the current simulation, clock drift is assumed to be $10^{-8}$ and the tolerance for being in synchronization is clocks that do not differ from actual time by more than $4 \times 10^{-8}$ seconds.

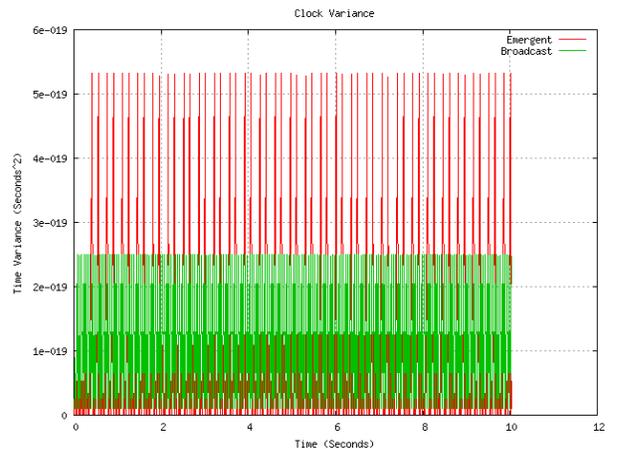

Fig. 6. The emergent technique shows larger variance in clock times, but not significantly worse than the broadcast technique.

## 4. THE IMPACT OF MOBILITY ON SYNCHRONIZATION

The simulation described in Section 3 has been enhanced with node mobility. The mobility model uses random motion with repulsive and attractive forces. Details of the movement forces are less important than the overall change in network node density. Nodes have been simulated with fast movement in order to accentuate the trends in the simulation. Movement characteristics were designed to be equivalent for broadcast time synchronization and emergent synchronization in order to provide a consistent measure of performance. As nodes move, node density changes. Node density is plotted in Fig. 7. Movement in this model is similar the spread of sensors in a fluid media.

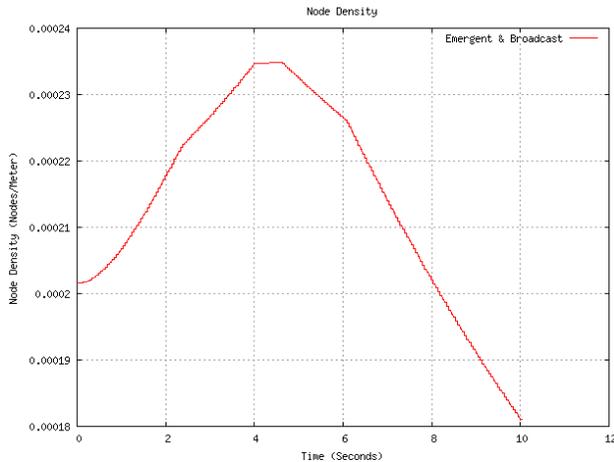

Fig. 7. Change in node density caused by node movement. Both simulations show similar decreases in density.

The synchronization power efficiency was measured to be approximately 12 times higher for the emergent technique. The emergent synchronization technique maintains a steady variance in clock times, shown in Fig.8, while the broadcast technique rises sharply.

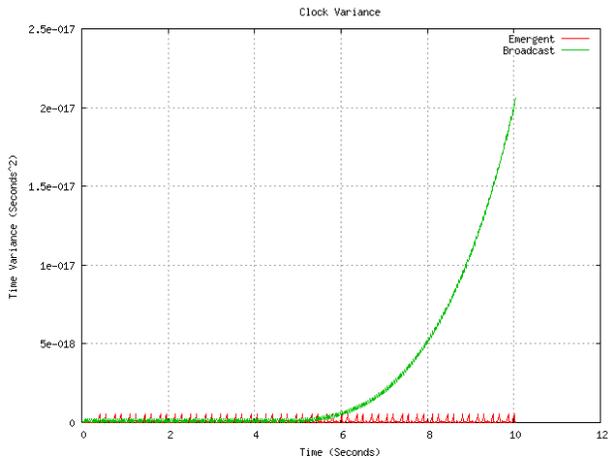

Fig.8. Clock variance shows a sudden increase with node mobility for the broadcast technique while having little perceptible effect on the emergent technique.

## 5. CONCLUSION

This paper has demonstrated that it is possible for rapid local exchanges of very short packet exchanges localized between neighboring nodes to achieve synchronized behavior with lower energy than centralized broadcast techniques. The emergent pulse coupled oscillation technique shows greater energy efficiency as well as robustness with mobility. Emergent techniques such as PCO need further research in order to explore their potential for further energy efficiency.

ACKNOWLEDGMENT

S.F. Bush thanks Anna Scaglione (Cornell University) for discussions on her prior work on this topic as well as Albert Davis and Carl Hein (Lockheed Martin Advanced Technology Laboratories) for their tireless effort in implementing an accurate CSIM model that provided the framework for the simulations in this paper.